\documentclass[sigconf]{acmart}
\AtBeginDocument{%
  \providecommand\BibTeX{{%
    \normalfont B\kern-0.5em{\scshape i\kern-0.25em b}\kern-0.8em\TeX}}}

\copyrightyear{2022}
\acmYear{2022}
\setcopyright{rightsretained}
\acmConference[IUI '22 Companion]{27th International Conference on Intelligent User Interfaces}{March 22--25, 2022}{Helsinki, Finland}
\acmBooktitle{27th International Conference on Intelligent User Interfaces (IUI '22 Companion), March 22--25, 2022, Helsinki, Finland}\acmDOI{10.1145/3490100.3516471}
\acmISBN{978-1-4503-9145-0/22/03}

\usepackage[capitalize]{cleveref}

\newcommand{\system}{SummaryLens}

\begin{document}

\title{\system{} -- A Smartphone App for Exploring Interactive Use of Automated Text Summarization in Everyday Life}

\renewcommand{\shorttitle}{\system{} -- An App for Exploring Interactive Text Summarization in Everyday Life}

\author{Karim Benharrak}
\email{karim.benharrak@uni-bayreuth.de}
\affiliation{%
  \institution{Department of Computer Science, University of Bayreuth}
  \city{Bayreuth}
  \country{Germany}
}

\author{Florian Lehmann}
\email{florian.lehmann@uni-bayreuth.de}
\affiliation{%
  \institution{Department of Computer Science, University of Bayreuth}
  \city{Bayreuth}
  \country{Germany}
}

\author{Hai Dang}
\email{hai.dang@uni-bayreuth.de}
\affiliation{%
  \institution{Department of Computer Science, University of Bayreuth}
  \city{Bayreuth}
  \country{Germany}
}

\author{Daniel Buschek}
\orcid{0000-0002-0013-715X}
\email{daniel.buschek@uni-bayreuth.de}
\affiliation{%
  \institution{Department of Computer Science, University of Bayreuth}
  \city{Bayreuth}
  \country{Germany}
}

\renewcommand{\shortauthors}{Benharrak et al.}

\begin{abstract}
  We present \system, a concept and prototype for a mobile tool that leverages automated text summarization to enable users to quickly scan and summarize physical text documents. We further combine this with a text-to-speech system to read out the summary on demand. With this concept, we propose and explore a concrete application case of bringing ongoing progress in AI and Natural Language Processing to a broad audience with interactive use cases in everyday life. Based on our implemented features, we describe a set of potential usage scenarios and benefits, including support for low-vision, low-literate and dyslexic users. A first usability study shows that the interactive use of automated text summarization in everyday life has noteworthy potential. We make the prototype available as an open-source project to facilitate further research on such tools.
\end{abstract}



\begin{teaserfigure}
    \centering
  \includegraphics[width=\textwidth]{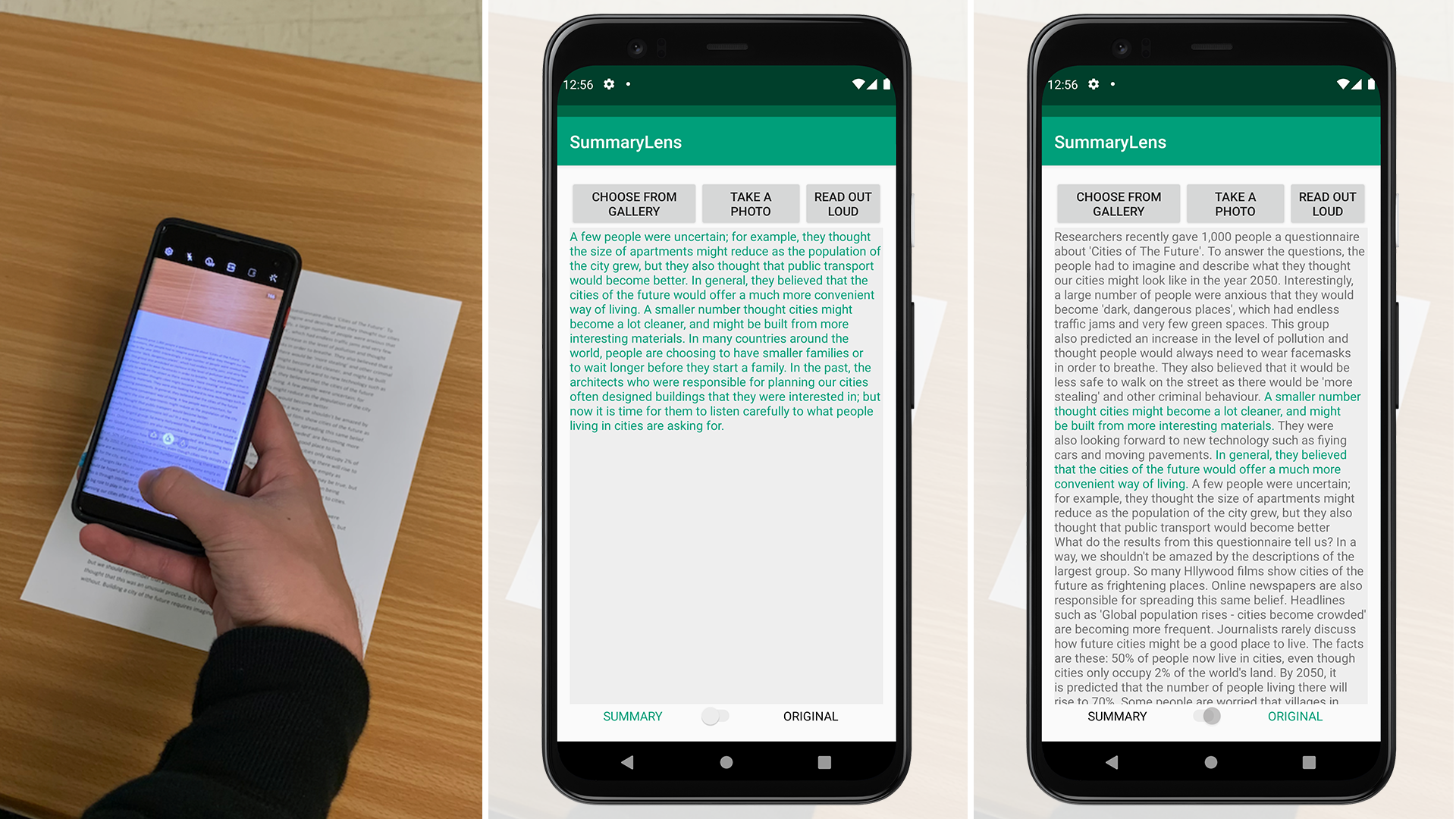}
  \caption{Our \textit{\system} prototype, implemented as an Android app that allows users to 1) take a photo of a physical document, 2) view an automatically created summary, and 3) use text-to-speech to read it out loud.}
  \Description{Three images: Hand holding smartphone taking a picture of a sheet of paper. Screenshot of app showing summarized text. Second screenshot showing full text with sentences that are also part of the summary highlighted in green.}
  \label{fig:teaser}
\end{teaserfigure}

\maketitle

\section{Introduction}

In today's corporate as well as private contexts it is essential to efficiently process text. Often, one would like to extract and understand the key aspects of a document, both to save time and to filter out (currently) irrelevant information. This is especially challenging for users who find prolonged focused reading difficult or impossible. Despite the many digital formats, we still also read and work with printed reports, articles, letters, newspapers, or other text documents. Extracting information is particularly challenging for such physical documents because people cannot rely, for example, on digital search or accessibility tools.
To address these challenges, we investigate the following guiding research question: \textit{How might we design a mobile tool to assist people in quickly comprehending physical documents in everyday tasks and environments?} %
Concretely, here we present our exploration of a solution strategy focused on enabling people to apply automatic text summarization to physical text documents. The result is \textit{\system}, a concept for a mobile application capable of scanning and summarizing text in real world environments (\cref{fig:teaser}). We implemented this concept as an Android app and report on a first user study. Our prototype is available as an open source project. %

\section{Related Work}

Following \citet{Radev2002}, the ``main goal of a summary is to present the main ideas in a document in less space''. We explore the case where this reduction further involves a transfer from the physical into the digital space.

Related work on interactive systems with text summarization mainly used it for digital text content: For example, \citet{Leiva2018} summarized text as part of content-responsive websites. Related, \citet{terHoeve2020} explored a conversational concept (i.e. chatbot, voice assistant) to help users extract information from a digital document. Moreover, \citet{Wang2021} proposed a system to create textual summaries of graphical user interfaces. Other work involved users to interactively improve AI text summaries~\cite{Gehrmann2020}. In contrast to the related work, we explore how people might be enabled to benefit from automatic summarization of \textit{physical} documents.

Technically, automated text summarization is an important task in ongoing work in Natural Language Processing~\cite{Kryscinski2019, Radford2019gpt2}. The two main approaches and types of summaries are \textit{extractive} (a selection of parts of the full text) and \textit{abstractive} (short rewrite of the full text)~\cite{Allahyari2017}. Our prototype supports both approaches in general.

\section{The \textit{\system} Concept and Prototype App}

\subsection{Concept: Combining Smartphone Camera, Summarization Model and Text-to-Speech System}
We propose the \textit{\system} concept as a concrete example application for bringing natural language processing capabilities to new everyday use cases.
Overall, with the described features, we aim to support diverse users in everyday life tasks by improving their experience and efficiency in understanding physical text documents.
Concretely, our concept has three parts, which map to steps of the corresponding user flow and interface.

\subsubsection{Scanning a physical text document}
In the first part of our concept, users ``scan'' a physical document with their mobile device camera (\cref{fig:teaser} left). %

\subsubsection{Summarizing the text}
After that, the text is presented in two modes in a graphical user interface:
The ``SUMMARY'' mode displays a summarized version of the scanned text (\cref{fig:teaser} center). 
While our concept supports various summarization methods, the concrete design discussed here targets extractive summarization on sentence level. That is, the summary shows the k most important sentences from the original text.
In contrast, in the ``ORIGINAL'' mode, the user can read the whole scanned text  (\cref{fig:teaser} right). In addition, %
sentences that are also included in the summary are highlighted in color. This makes it transparent to the user which sentences are considered essential by the system in the context of the entire text.
The user can switch back and forth between both modes.

\subsubsection{Reading out the text}
Finally, our concept includes a ``READ OUT LOUD'' functionality, which utilises a Text2Speech service to enable the user to listen to the currently displayed text (original or summary). %

\subsection{Prototype Implementation as an Android App}
We developed a \textit{\system} prototype as an Android application and a Python backend and server. The app uses Optical Character Recognition (OCR)\footnote{see \url{https://developers.google.com/ml-kit/vision/text-recognition}, last accessed 03.02.2022} to recognize and extract text from a photo captured with the smartphone camera. This extracted text is saved and sent to our Python API (developed using Flask) for summarization. Concretely, the sentences within the text are ranked by relevance using a text ranking algorithm and typical preprocessing steps. In particular, our default uses GloVe embeddings and TextRank~\cite{pennington2014glove, mihalcea2004textrank} to identify the top k sentences.

Our backend returns the top five sentences to the Android app in order of their appearance in the original text. This summary is saved and shown to the user. The user may switch between viewing the summary and the full text using a toggle button (see bottom of the UI in \cref{fig:teaser}). For extractive summaries, our app implements the described highlighting concept: When viewing the full text in this way, the top sentences are colored in green. 

In addition, the app integrates Android's text-to-speech API and exposes this functionality via a button (see top right of the UI in \cref{fig:teaser}). This is currently limited to reading in English, but could be flexibly extended. Users can cancel reading the text by pressing the button again.

We consider this prototype a proof of concept with the intention of serving as a practical code base and starting template for further research. Our implementation is flexible and can be extended, for example, with other summarization methods in Python. For instance, we have also implemented frequency-based summaries and (abstractive) summarization with Deep Learning models, using \textit{HuggingFace}~\cite{wolf_huggingfaces_2020}.

\section{User Study}

To gain first insights into users' experience and views with our concept and prototype, we conducted an exploratory user study with five students (mean age 22 years, range 20-28). 

We installed our application on a study smartphone and prepared a table with a printed text document. Since it was not our focus to evaluate the quality of the OCR API we used, we simulated ``perfect'' OCR in the study by storing the text of this document in the app directly. The app can then create the summary based on the stored text, using the described summarization backend.

Participants were informed about the content and purpose of the study and signed a consent form. We then instructed them to try out the app, involving the text document, while thinking aloud. No further explanation of the UI was provided. 
We observed the process of people exploring the prototype and took notes of these observations and participants' comments and shared experiences. %
After they had completed the task (i.e. scanned the document and explored all app functionalities), we conducted a short semi-structured interview around three central questions: \textit{What did you like the most and why?}, \textit{What do you wish would be different and what needs to be improved?}, \textit{In which everyday situations would you like to have an application like that and why?}

\section{Results and Discussion}

People's comments indicated two potential key benefits: First, creating summaries on the go allows for faster document reading, which is useful in time-sensitive scenarios. Second, it might also facilitate comprehension (e.g. gaining an overview, extracting key points). %
Saving time was mentioned by everyone (e.g.: \textit{``On the one hand you have your full text and then just with one click you can access a summary which is very convenient because it saves a lot of time.''}). To realise this, the ability to photograph a document or select an image from the phone's gallery was emphasised as highly important. These comments and observations provide promising feedback on our fundamental concept.

Regarding our prototype's UI, people found the ability to switch between the original text and the related summary easy and fast to use. %
This feature was also seen as making it easier to go deeper into the full text as needed. %
From our observations and people's comments, the sentences highlighted in green in the original text here served as landmarks for orientation, facilitating more selective reading in the full text. People further commented on the straightforward layout and easily recognizable steps of the user flow from picture to summary in particular.

We also identified areas for improvement and extension: People wished to be able to alter the font size, to adjust the voice's reading speed, to include support for PDF formats and handwritten text, and summarisation settings (e.g. \textit{``I would like to have a longer summary''}).
We also received input on explainability: Some wondered if the offered summary did not omit vital information, and were curious about the used summarization method (\textit{``I thought it was interesting how they summarized the points, however I don't know how accurate the summary is.''}). From this, we learn that the concept and prototype should convey to the user how parts or aspects of the full text are being chosen for the summary. This would facilitate understanding of the process and potentially increase trust. For our extractive method here, the highlighting in the full text can serve as a first step but should be explored further. 

Finally, the study also revealed application ideas, such as reading a physical newspaper article in a hurry and storing it to make it available on the go. Another idea mentioned was catching up with longer documents at work, potentially including multitasking situations (e.g. gaining and maintaining an overview when working across multiple documents).

\section{Conclusion}

We have presented \textit{\system}, a concept and prototype for a mobile tool that leverages automated text summarization to enable users to quickly summarize physical text documents. In the larger context of our research, this explores a concrete example of an interactive tool aimed at rendering ongoing progress in AI and Natural Language Processing useful to people in everyday life situations. We plan to refine and extend this concept and prototype based on the insights from our first user study. To facilitate further work in this direction more broadly, we also release our prototype app and backend as an open-source project to the community here: 
\url{https://github.com/DerKarim06/SummaryLens}

\begin{acks}
This project is funded by the Bavarian State Ministry of Science and the Arts and coordinated by the Bavarian Research Institute for Digital Transformation (bidt).
\end{acks}

\bibliographystyle{ACM-Reference-Format}
\bibliography{bibliography}

\end{document}